\documentclass[%
 amsmath,amssymb,aps,floatfix
]{revtex4-1}

\usepackage[dvipdfmx]{graphicx}
\usepackage{dcolumn}
\usepackage{bm}
\usepackage{hyperref}

\newcommand{\MIN}{\mathop{\rm min}\limits}
\newcommand{\LONGARROW}{\mathop{\longrightarrow}\limits}
\newcommand{\figwidth}{0.49\textwidth}

\graphicspath{ {./} } 

\begin{document}

\title[Transition properties in dynamical and statistical features of drying crack patterns]{Transition properties in dynamical and statistical features of drying crack patterns}
\author{Shin-ichi Ito}
\affiliation{Earthquake Research Institute, The University of Tokyo, 1-1-1 Yayoi, Bunkyo-ku, 113-0032, Tokyo, Japan}
\affiliation{Department of Mathematical Informatics, Graduate School of Information Science and Technology, The University of Tokyo, 7-3-1, Hongo, Bunkyo-ku, 113-8656, Tokyo, Japan}

\author{Akio Nakahara}
\affiliation{ Laboratory of Physics, College of Science and Technology, Nihon University, 7-24-1 Narashinodai, Funabashi, 274-8501, Chiba, Japan}

\author{Satoshi Yukawa}
\affiliation{Department of Earth and Space Science, Graduate School of Science, Osaka University, 1-1 Machikaneyama-cho, Toyonaka, 560-0043, Osaka, Japan}

\date{\today}
\begin{abstract}
In this study, we experimentally investigated the time dependence of the statistical properties of two-dimensional drying crack patterns to determine the functional form of fragment size distribution.
Experiments using a thin layer of a magnesium carbonate hydroxide paste revealed a ``dynamical scaling'' property in the time series of the fragment size distribution, which has been predicted by theoretical and numerical studies.
Further analysis results based on Bayesian inference show the transition of the functional form of the fragment size distribution from a log-normal distribution to a generalized gamma distribution.
The combination of a statistical model of the fragmentation process and the dynamics of stress concentration of a drying thin layer of viscoelastic material explains the origin of the transition.
\end{abstract}

\maketitle

\section{Introduction\label{level.1}}
Investigating the statistical features of fragments in crack patterns is crucial to understanding the physical properties of objective materials, and the history of fracturing events that such materials experience. 
The size distribution of the fragments in crack patterns is a fundamental and vital statistical feature because it exhibits universality in its functional form. 
For instance, it is known that the size distribution of the fragments in brittle materials takes two functional forms: a log-normal distribution and a power-law distribution~\cite{ishii1992, Oddershede93, kadono1997}.

Surface crack patterns appear when dense colloidal suspensions (pastes) are dried.
Such ``drying crack'' patterns can be observed in dried lakes, paddy fields, and paintings.
Cell-like or network-like patterns are the most common types of patterns in drying cracks. However, physically interesting patterns may sometimes form, owing to nonlinear rheological properties~\cite{Nakahara05, Nakahara06, Matsuo12}.
It is notable that drying cracks have a remarkable similarity to the morphological features observed in the crack patterns in cooling lava. Owing to this observation, drying cracks are used in surrogate experiments to investigate the formation of crack patterns such as columnar joints in cooling lava~\cite{MULLER98a, MULLER98b, Hamada20} to understand the physical origin of the crack pattern formation of the cooling lava. Drying cracks grow slowly, depending on the rate of evaporation of the inner liquid ~\cite{Kitsunezaki09}.
The resulting fragment size distribution depends not only on the material properties of pastes but also on the history of the drying process.
This phenomenon leads to the expectation that the functional form of the fragment size distribution holds the information regarding the drying history.
However, there is limited experimental knowledge on the time evolution of the fragment size distribution, and even the functional form of the fragment size distribution of completely-dried-out crack patterns is not well established~\cite{lecocq2002}, because obtaining a sufficient number of samples to construct an accurate histogram of the fragment size distribution is difficult in experiments.

In this study, we investigated the time-dependence of the fragment size distribution in the drying crack patterns of a thin layer of paste, and extracted useful information to analyze the drying history from the functional form of the fragment size distribution experimentally.
Constructing an accurate histogram of the fragment size distribution requires a vast number of fragments. The number of fragments obtained from experiments, especially during the early stages of fragmentation, is not sufficiently large.
Instead of constructing a histogram, we consider several functional forms to estimate the fragment size distribution accurately.
Bayesian inference~\cite{box1973bayesian} plays an essential role in the assessments of the assumed functional forms.
Bayesian inference enables the acquisition of valuable information to determine the functional form from a smaller number of samples~\cite{Ito19}, and provides objective assessments of the assumed functional forms~\cite{schwarz1978estimating, Akaike1998, Nagata12}.

The remainder of this paper is organized as follows: 
In Section~II, we discuss the experimental setup and present the experimental fragment size distribution.
In Section~III, we present the analysis of the functional form of the fragment size distribution on the basis of Bayesian inference. In Section~IV, we discuss the origin of the functional form through the investigation of a statistical model of a fragmentation process and the dynamics of stress concentration in a drying thin layer of viscoelastic material. Finally, we present a concluding discussion in Section~V.

\section{Experiments\label{level.2}}
\begin{figure*}[tbp]
\centering
\includegraphics[width=0.95\textwidth]{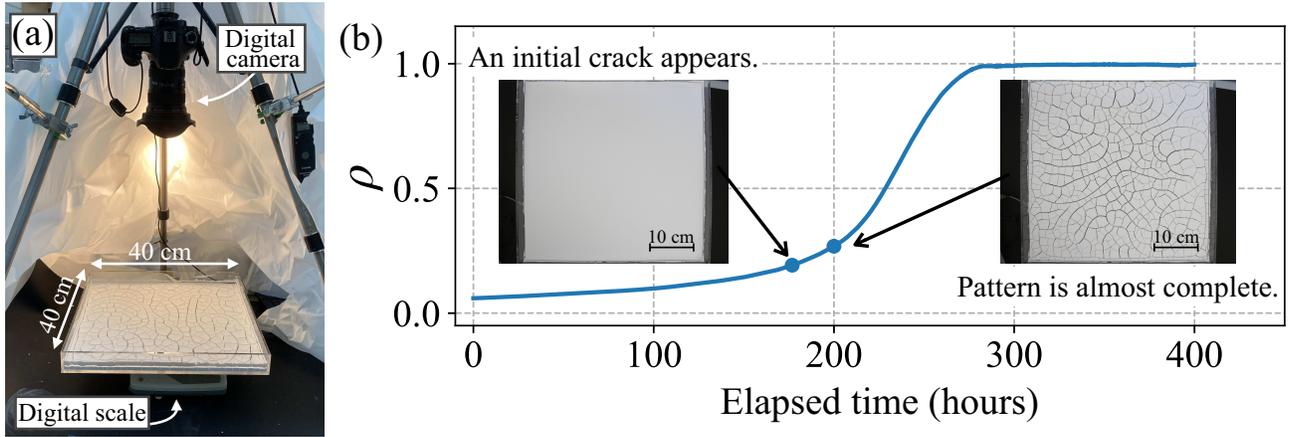}
\caption{(a) Experimental setup and (b) typical time evolution of the solid volume fraction $\rho$.\label{fig:experimet}}
\end{figure*}

In this section, we explain the experimental setup and analyze fragment data by building the size distribution.
Figure~\ref{fig:experimet}(a) shows the experimental setup.
We employed a mixture of distilled water and magnesium carbonate hydroxide powder manufactured by Kanto Chemical, Japan. The mass density of the powder was $2.0\;\text{g}\slash\text{cm}^{3}$.
The paste was dried in an acrylic resin container naturally at room environment (temperature $25{}^\circ\mathrm{C}\pm 1{}^\circ\mathrm{C}$ and humidity $10\%$--$50\%$), and a digital camera fixed above the container recorded the time evolution of the surface of the paste.
In addition, the digital scale placed under the container measured the changes in the mass of the paste over time.
The mass was used to calculate the time evolution of a solid volume fraction $\rho$, defined by
\begin{equation}\label{eq:2.0}
\rho = \frac{V_{\text{pow}}}{V_{\text{wat}}+V_{\text{pow}}},
\end{equation}
where $V_{\text{pow}}$ and $V_{\text{wat}}$ are the volumes of the powder and water in the container, respectively.
We set the initial solid volume fraction and the thickness of the paste to $6\%$ and $2\;\text{cm}$, respectively, at the start of the experiment, and then recorded the time evolution of the surface pattern and the solid volume fraction until the paste was completely dried out.
We repeated this procedure five times with different samples.
Figure~\ref{fig:experimet}(b) shows a typical time evolution of the solid volume fraction $\rho$.
The crack patterns evolve slowly over time, and it takes more than one day from the appearance of the initial crack to the end of the pattern evolution.
We can infer from this result that the stress field in the paste is sufficiently relaxed, and is purely driven by the negative pressure that depends only on $\rho$~\cite{Kitsunezaki09}. This pressure increases monotonically with time, and thus, the variation of the statistical properties of the patterns also depend only on $\rho$.
For this reason, we treated the variation of $\rho$ as an ``elapsed time'' variation. This enables the realization of high statistical accuracy, by merging the data from different samples at the same solid volume fraction.

\begin{figure}[tbp]
\centering
\includegraphics[width=\figwidth]{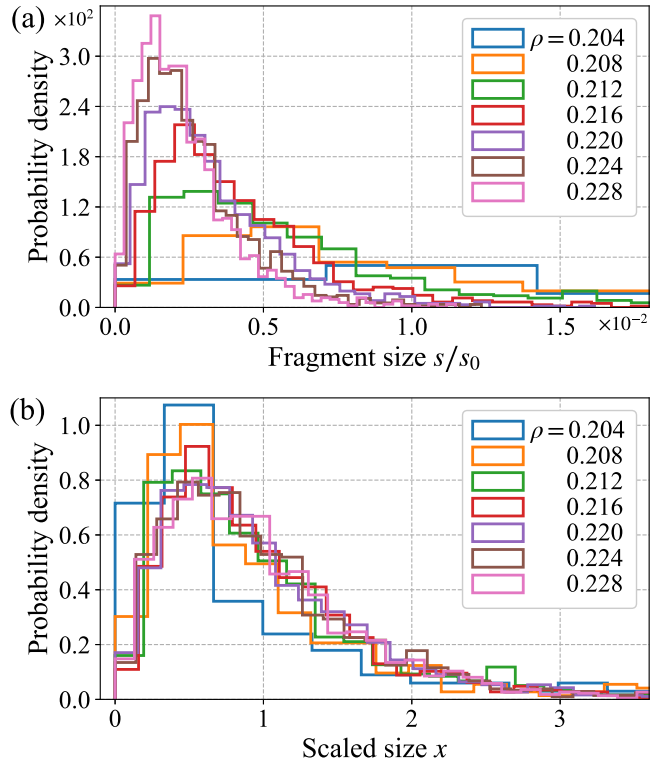}
\caption{Variation of the fragment size distribution as a function of $\rho$. (a) Raw fragment size distribution; the fragment sizes are scaled by the areal size of the top surface of the container as $s_{0} = 1,600\;\text{cm}^2$. (b) Fragment size distribution scaled by the average $\left<s\right>$, i.e., the horizontal axis indicates the scaled size $x = s\slash\left<s\right>$. \label{fig:pdfs}}
\end{figure}
We obtained fragment size distributions for each $\rho$ by extracting the upper surface areal size of the fragments from the obtained images of crack patterns~\footnote{We used the \textsf{OpenCV} Python package for image analysis.}.
Figure~\ref{fig:pdfs}(a) shows the fragment size distribution at each solid volume fraction.
It shows that the distribution is almost unimodal, and the mode shifts to smaller values as $\rho$ increases, meaning that the typical fragment size decreases monotonically with time.
The series of the fragment size distributions has an interesting property in that it can be scaled by the average fragment size, which varies with $\rho$, when the solid volume fraction becomes larger than approximately $0.216$, as shown in Fig.~\ref{fig:pdfs}(b).
This result implies that the time-dependent fragment size distribution $f_{t}(s)$ obeys an asymptotic form:
\begin{equation}\label{eq:2.1}
f_{t}(s)ds \LONGARROW_{t\rightarrow\infty} f(x)dx,
\end{equation}
where $t$ is the time, $s$ is the fragment size, $f(x)$ is a time-invariant probability density, and $x$ is the scaled fragment size given by
\begin{equation}\label{eq:2.2}
x = \frac{s}{\left<s\right>},
\end{equation}
where $\left<s\right>$ is the average fragment size defined by the arithmetic mean of the fragment size samples.
Although this ``dynamical scaling'' property has been reported through numerical simulations of drying crack patterns~\cite{ito2014-1,zoltan2017}, to the best of our knowledge, this is the first experimental confirmation of the process.

\section{Bayesian inference}
In this section, we discuss the details of the functional form of the scaled fragment size distribution $f(x)$, on the basis of Bayesian inference.
Our Bayesian inference will be used to evaluate three parametric models for the fragment size distribution based on dataset $\mathcal{D}$ of the scaled fragment size obtained at each $\rho$. Then, based on an information criterion, we will select the best model for the fragment size distribution that underlies the dataset.
The parametric models as candidates for the scaled fragment size distribution considered here are as follows:
\begin{itemize}
\item[(i)] Log-normal distribution \cite{lecocq2002, zoltan2017}
\begin{equation}\label{eq:lognorm}
f\left(x\mid\nu\right) = \dfrac{1}{\sqrt{2\pi\nu}x}\exp\left[-\frac{\left(\log x+\nu\slash 2\right)^2}{2\nu}\right],
\end{equation}
where $\nu$ is a positive parameter.
\item[(ii)] Weibull distribution \cite{lecocq2002}
\begin{equation}\label{eq:weibull}
f\left(x\mid m\right) = Am\left(Ax\right)^{m-1}\exp\left[-\left(Ax\right)^m\right],
\end{equation}
where $m$ is a positive parameter, and the scale parameter $A$ is chosen such that the expectation of $x$ is unity.
\item[(iii)] Generalized gamma distribution 
\begin{equation}\label{eq:gengamma}
f\left(x\mid d,g\right) = \frac{Bg}{\Gamma(d/g)}\left(Bx\right)^{d-1}\exp\left[-\left(Bx\right)^g\right],
\end{equation}
where $d$ and $g$ are positive parameters, and $\Gamma(z)$ is a gamma function. The scale parameter $B$ is chosen such that the expectation of $x$ is unity.
\end{itemize}
Models~(i) and (ii) have been employed as the candidates of the fragment size distribution of the drying crack patterns in previous studies~\footnote{Unlike in our work, the drying crack experimentally investigated in Ref.~\cite{lecocq2002} is constrained in one dimension.}.
Model~(iii) is a generalization of model~(ii), i.e., model~(iii) constrained to $d = g$ is equivalent to model~(ii).
For simplicity of notation, the parameters in each model are described as a parameter vector $\theta$, and each model is represented as $f(x\mid\theta)$.
The parametric estimation starts from building a posterior probability density $p(\theta\mid\mathcal{D})$ of the parameter vector $\theta$ with a given dataset $\mathcal{D}$, on the basis of Bayes' theorem
\begin{equation}\label{eq:2.3}
p(\theta\mid\mathcal{D}) = Cp(\theta)p(\mathcal{D}\mid\theta),
\end{equation}
where $C$ is a normalization constant, $p(\theta)$ is a prior probability density that includes a priori information of the parameter vector $\theta$, and $p(\mathcal{D}\mid\theta)$ is a likelihood function.
Assuming that the elements in the dataset $\mathcal{D}$ follow a model $f(x\mid\theta)$ identically and independently leads to the likelihood function 
\begin{equation}\label{eq:2.4}
p(\mathcal{D}\mid\theta) = \prod_{x\in\mathcal{D}}f(x\mid\theta).
\end{equation}
Additionally, because there is no information related to the elements in $\theta$ except for the positivity, we employ a prior probability density given by an exponential density as
\begin{equation}\label{eq:2.5}
p(\theta\mid\eta) = \left(\prod_{\eta_{i}\in\eta} \eta_{i}\right) \exp(-\eta^{\top}\theta),
\end{equation}
where the vector $\eta$ is a hyper-parameter vector that determines the broadness of the prior probability density and is to be optimized later. $\bullet^{\top}$ indicates the transpose of $\bullet$, and we rewrite the prior probability density from $p(\theta)$ to $p(\theta\mid\eta)$ to describe its $\eta$-dependency explicitly.
The combination of Eqs.~\eqref{eq:2.3}--\eqref{eq:2.5} with the given dataset $\mathcal{D}$ and the hyper-parameter vector $\eta$ yields the posterior probability density.

\begin{figure}[tbp]
\centering
\includegraphics[width=\figwidth]{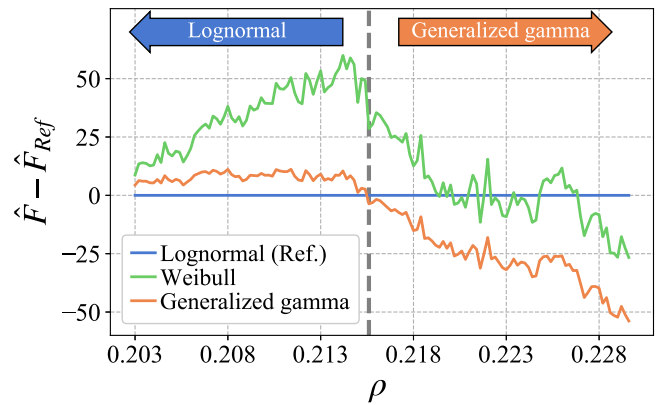}
\caption{Free energy as a function of $\rho$. Each solid line shows the difference between the free energy of each model and that of the log-normal.
The vertical dashed line indicates the solid volume fraction where the best model switches from the log-normal distribution to the generalized gamma distribution.\label{fig:fe}}
\end{figure}

\begin{figure}[tbp]
\centering
\includegraphics[width=\figwidth]{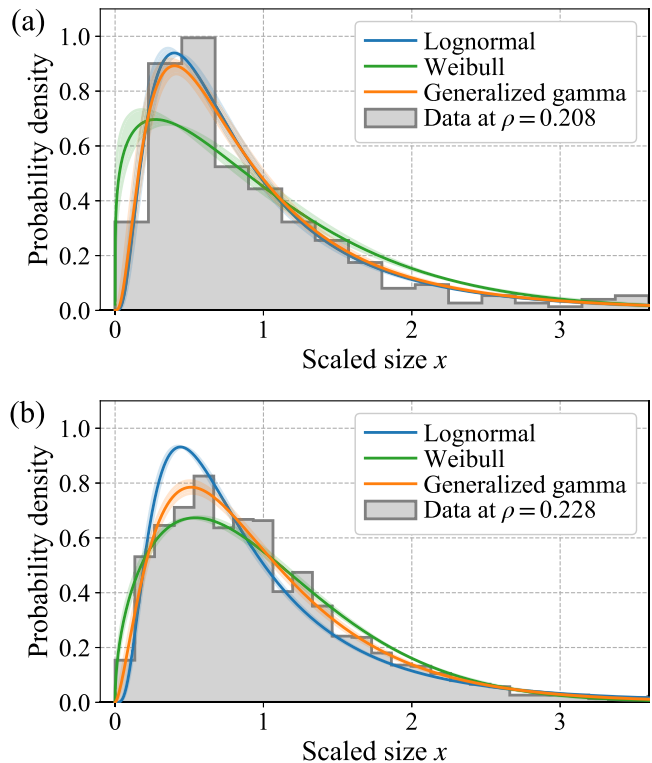}
\caption{Comparison of the histogram data and fragment size distribution predicted by our Bayesian inference. The transparent region around each line denoting the probability density indicates the $90\%$ credible interval. In (a) and (b), the datasets at $\rho = 0.208$ and $0.228$ are used, respectively. \label{fig:predict}} 
\end{figure}

An advantage of Bayesian inference is that it enables the evaluation of the ``goodness of the model'' on the basis of information the posterior probability density contains~\cite{schwarz1978estimating, Akaike1998}.
Taking into consideration that the dataset $\mathcal{D}$ at each solid volume fraction, especially the dataset at the early stage of the fragmentation process, may have an insufficient data size that does not guarantee the Gaussianity of the resulting posterior probability density, we employ the ``free energy'' defined by
\begin{equation}\label{eq:2.6}
F(\eta) = -\log\int d\theta\; p(\theta\mid\eta)p(\mathcal{D}\mid\theta),
\end{equation}
to measure the goodness of model~\cite{Nagata12}. 
The integral included in the free energy describes a conditional probability of the dataset given the pair $p(\theta\mid\eta)$ and $f(x\mid\theta)$, meaning that the better pair yields the smaller free energy.
We find the best model that minimizes the free energy among models~(i)--(iii) with the given fragment size dataset at each $\rho$.
We compute the free energy optimized with respect to $\eta$ as
\begin{equation}\label{eq:2.6-1}
\hat{F} = \MIN_{\eta}F(\eta),
\end{equation}
for each model~\footnote{We used the \textsf{GPyOpt} Python package for the optimization.}, and then compare them at $\rho$.
Figure~\ref{fig:fe} shows the optimized free energy as a function of $\rho$.
These observations state that the Weibull distribution~(ii) is always rejected, and the best model switches from the log-normal distribution~(i) to the generalized gamma distribution~(iii) at $\rho = 0.215$--$0.216$.
It is notable that this transition point overlaps the point where the dynamical scaling begins to appear, as seen in Fig.~\ref{fig:pdfs}(b).
This result suggests that the fragment size distribution transits from a log-normal distribution that does not have the dynamical scaling property, to a generalized gamma distribution that has the scaling property.
Figure~\ref{fig:predict} shows the fragment size distribution weighted by the posterior probability density
\begin{equation}\label{eq:2.7}
f^{*}(x\mid\mathcal{D}) = \int d\theta\; f(x\mid\theta) p(\theta\mid\mathcal{D}),
\end{equation}
in a $90\%$ credible interval.
Figures~\ref{fig:predict}(a) and (b) respectively show the fragment size distributions before and after the transition.
This result visually confirms that the log-normal distribution appears to be better than the other models in the early stage before the transition point, although the credible interval is slightly large because of the small data size, and the generalized gamma distribution is better in the later stage.

\section{Theory\label{level.3}}
The experiments and analysis based on Bayesian inference demonstrated a transition of the functional form of the scaled fragment size distribution from a log-normal distribution to a generalized gamma distribution.
The following theory based on Ref.~\cite{ito2014-2} provides an explanation of the transition.
The time-dependent fragment size distribution $f_{t}(s)$ obeys a master equation
\begin{equation}\label{eq3.1}
\frac{\partial f_{t}(s)}{\partial t} = -\lambda_{s} f_{t}(s) + \int_{0}^{\infty} ds' w_{s'\rightarrow s} \lambda_{s'}f_{t}(s'),
\end{equation}
where $\lambda_{s}$ is an intensity function depending on $s$.
The function $w_{s'\rightarrow s}$ is a transition probability from $s'$ to $s$, defined by
\begin{equation}\label{eq3.1-1}
w_{s'\rightarrow s} = \int_{0}^{1}dr\; \delta(rs'-s)q(r) = \frac{1}{s'}q\left(\frac{s}{s'}\right),
\end{equation}
where $\delta(z)$ indicates the Dirac delta function, and $q(r)$ is a probability density of the ratio $r\in[0,1]$ of $s$ to $s'$.
The discussion in Appendix~\ref{AppA1} shows that Eq.~\eqref{eq3.1} with a constant intensity function leads to the fragment size distribution $f_{t}(s)$, converging to a log-normal distribution that does not have the dynamical scaling property.
Moreover, the fragment size distribution exhibits the scaling property theoretically if and only if the intensity function is a power function of $s$, i.e., $\lambda_{s}\propto s^{\gamma}$ for a nonzero $\gamma$~\cite{ito2014-2}.
In this case, the tail of the scaled fragment size distribution $f(x)$ obeys $\exp\left[-\left(x\slash c\right)^{\gamma}\right]$, where $c$ is a constant, and the functional form in the neighborhood of the origin is characterized by a power function $x^{\alpha}$ whose exponent $\alpha$ depends on the functional form of $q(r)$, i.e.,
\begin{equation}\label{eq3.2}
f(x)\sim\left\{
\begin{array}{ccl}
x^{\alpha} & \text{for} & x\sim 0 \\
\exp\left[-\left(x\slash c\right)^{\gamma}\right] & \text{for} & x\gg 1.
\end{array}\right.
\end{equation}
Asymptotically, this essentially is a generalized gamma distribution.
Appendix~\ref{AppA2} contains details.
These theoretical facts suggest that the intensity function transits from a constant function to a power function in a fragmentation process of drying crack patterns.
Because the reciprocal of the intensity function is a characteristic decaying time scale of the existence probability of fragments having a size $s$, the transition of the intensity function is explained by the fragment-size dependency of the decaying time scale.
Here, in order to discuss the fragment-size dependency of the decaying time scale, we consider the drying process of a thin layer of viscoelastic material that adheres to a flat substrate.
The balance equation and the two-dimensional stress equation are 
\begin{equation}\label{eq3.3}
\begin{aligned}
\nabla\cdot\sigma & = ku \\
\sigma & = \mathsf{C}:\left\{\nabla u +(\nabla u)^\top\right\} + h(t)\mathsf{I},
\end{aligned}
\end{equation}
where $u$ and $\sigma$ are a two-dimensional displacement field and a stress tensor field, respectively, and $\mathsf{C}$ is a constant elastic coefficient tensor.
The term $ku$ describes the resistance force arising from the adhesion of the material to the substrate, where $k$ is a constant.
The diagonal tensor $h(t)\mathsf{I}$ indicates a negative pressure that increases with time, where $h(t)$ is a monotonically increasing scalar function of time, and $\mathsf{I}$ is a two-dimensional unit tensor. 
Although obtaining an exact solution of Eq.~\eqref{eq3.3} with a given boundary condition (i.e., shape of fragment) is generally difficult, evaluating the characteristic scales involved in Eq.~\eqref{eq3.3} enables us to estimate the behavior of the characteristic stress that appears in the fragment.
Let $U$ and $S$ be the characteristic scales of the displacement and stress, respectively.
When the shape of the fragment is less complex, the characteristic length scale involved in $u$ and $\sigma$ can be proportional to $L$, which is a square root of the areal size $s$ of the fragment.
Replacing the quantities in Eq.~\eqref{eq3.3} with the characteristic scales yields
\begin{equation}\label{eq3.4}
\begin{aligned}
-\frac{S}{L} & = kU \\
S & = E\frac{U}{L} + h(t),
\end{aligned}
\end{equation}
where $E$ is a characteristic elastic coefficient. Then, eliminating $U$ from Eq.~\eqref{eq3.4} yields
\begin{equation}\label{eq3.5}
S = \frac{h(t)}{1+\lambda_{D}^{2}\slash s},
\end{equation}
where $\lambda_{D}=\sqrt{E\slash k}$ is the characteristic length scale determined by the material constants of the fragment and substrate.
Because $h(t)$ is an increasing function of time, the characteristic stress $S$ also increases with time.
Assuming that the fragment breaks when the stress $S$ approaches a certain threshold stress $\sigma_{Y}$, the time scale $T$ required until the fragment breaks is estimated by
\begin{equation}\label{eq3.6}
T = h^{-1}\left(\sigma_{Y}\left( 1+ \frac{\lambda_{D}^{2}}{s}\right)\right),
\end{equation}
where $h^{-1}$ indicates the inverse function of $h$.
Equation~\eqref{eq3.6} shows that the time scale $T$ has two asymptotic behaviors depending on the fragment size $s$:
\begin{equation}\label{eq3.7}
T = \left\{
\begin{array}{ccl}
h^{-1}\left(\sigma_{Y}\right)=\text{const.} & \text{for} & s\gg\lambda_{D}^2 \\
h^{-1}\left(\sigma_{Y}\dfrac{\lambda_{D}^{2}}{s}\right) & \text{for} & s\ll\lambda_{D}^2.
\end{array}\right.
\end{equation}
The asymptotic forms of $T$ in Eq.~\eqref{eq3.7} suggest that the fragments larger than $\lambda_{D}^2$ break within a constant time scale, and then as the fragmentation process continues, fragments smaller than $\lambda_{D}^2$ begin to appear. These break with the size-dependent time scale.
The constant time scale in the early stage of the fragmentation process of drying crack patterns provides the fragment size distribution of the log-normal form, and the size-dependent time scale in the later stage suggests that the fragment size distribution does not obey the log-normal form.
Because dynamic scaling is realized if and only if the intensity function $\lambda_{s}(\sim 1/T)$ is given by a power function of $s$, the negative pressure $h(t)$ is also a power function of time.
Although directly measuring $h(t)$ in actual experiments is a future research problem, the fact that the dynamical scaling property is observed strongly suggests that $h(t)$ is a power function of time.

\section{Conclusions}
We investigated the time evolution of the size distribution of fragments in two-dimensional drying crack patterns, and confirmed the scaling property predicted numerically in previous studies.
Our Bayesian inference based on the free energy revealed a dynamic transition of the functional form of the fragment size distribution.
The origin of the transition is explainable from the combination of a statistical model of a fragmentation process and the dynamics of stress concentration in a drying thin layer of viscoelastic material.

According to our theoretical results, the tail of the scaled fragment size distribution is determined by the exponent of the negative pressure.
This result suggests that investigating the exponent of the tail enables us to estimate the negative pressure related to the history of the drying process from a completely-dried-out crack pattern, even if the time evolution of the crack pattern is not available.

Although we chose magnesium carbonate hydroxide from the perspective of ease of fragment detectability, investigating whether other powders exhibit similar results is an interesting topic of further research. 
In practice, some kinds of powders exhibit patterns in which the connection of the cracks is ambiguous and undetectable, unlike the cell patterns obtained in this study.
Such cases require designing other types of characteristic quantities that are not fragment-based.

Investigating the time-dependent property of the fragment size distribution in drying three-dimensional materials is also appealing. The functional form may hold the information of unmeasurable or difficult-to-measure quantities, such as the spatial distribution of moisture contents~\cite{Mizuguchi05}.
To the best of our knowledge, the detailed functional form of the fragment size distribution in the three-dimensional case has not been reported numerically or experimentally, and remains as an open problem.

\begin{acknowledgments}
This work was mainly supported by a JSPS Grant-in-Aid for Young Scientists (B), Grant Number JP19K14671. The key technique in this work was triggered by discussions in the research projects of JSPS Grant-in-Aid for Scientific Research (C), Grant Number JP19K03652, and JST CREST Grant Numbers JPMJCR1761 and JPMJCR1763. The authors would like to thank Editage (www.editage.com) for English language editing.
\end{acknowledgments}

\appendix
\section{Solution of the master equation}\label{AppA}
This appendix presents the solutions of the master equation (Eq.~\eqref{eq3.1}) for two cases of the functional form of the intensity function $\lambda_{s}$: One is the case of a constant, and the other is the case of a power function of $s$.
The initial condition of the fragment size distribution is assumed here to be $f_{0}(s) = \delta(s-\tilde{s})$, where $\tilde{s}$ is the initial fragment size.
\subsection{Case of a constant intensity function}\label{AppA1}
Under the assumption that the intensity function is $\lambda_{s} = 1\slash\tau$, where $\tau$ is a constant time, the master equation can be expressed as 
\begin{equation}\label{eqA.1}
\tau \frac{\partial f_{t}(s)}{\partial t} = -f_{t}(s) + \int_{s}^{\infty} \frac{ds'}{s'}q\left(\frac{s}{s'}\right) f_{t}(s'),
\end{equation}
This equation has an explicit solution 
\begin{equation}\label{eqA.2}
f_{t}(s) = \frac{1}{2\pi s}\int_{-\infty}^{\infty}dk\; \exp\left[ik\log\frac{s}{\tilde{s}}+\frac{t}{\tau}\left\{w(k)-1\right\}\right],
\end{equation}
where 
\begin{equation}\label{eqA.3}
w(k) = \int_{0}^{1}dr\; q(r)\exp\left(-ik\log r\right).
\end{equation}
When $t\rightarrow\infty$, this solution asymptotically follows a log-normal distribution of the form:
\begin{equation}\label{eqA.4}
f_{t}(s) = \frac{1}{\sqrt{2\pi (t\slash\tau)m_{2}}s} \exp\left[-\frac{\left\{\log\left(s\slash\tilde{s}\right)-(t\slash\tau)m_{1}\right\}^2}{2(t\slash\tau)m_{2}}\right],
\end{equation}
where the constants $m_{l}$ $(l=1,2)$ are given by
\begin{equation}\label{eqA.5}
m_{l} = \int_{0}^{1}dr\;q(r)\left(\log r\right)^{l}.
\end{equation}
It is easily checked that the asymptotic solution (Eq.~\eqref{eqA.4}) remains time-dependent, even when scaling the fragment size by its expectation 
\begin{equation}\label{eqA.6}
\left<s\right> = \tilde{s}\exp\left[\frac{t}{\tau}\left(m_{1}+\frac{m_{2}}{2}\right)\right].
\end{equation}
This means that the master equation with a constant intensity function provides a fragment size distribution covering a log-normal distribution that does not have the dynamical scaling property (Eq.~\eqref{eq:2.1}).
\subsection{Case of a power intensity function}\label{AppA2}
\begin{figure}[tbp]
\centering
\includegraphics[width=\figwidth]{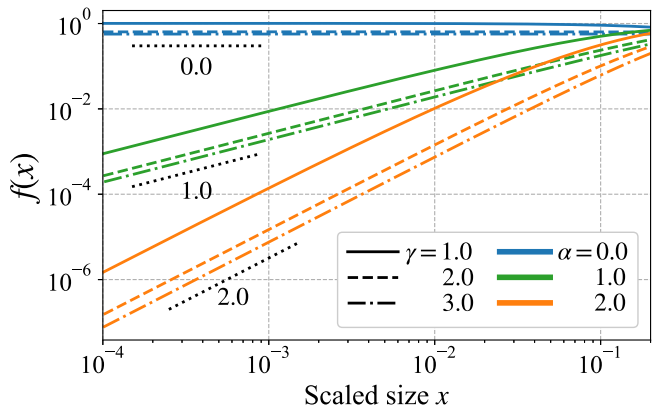}
\caption{Numerical solutions of the scaled master equation (Eq.~\eqref{eqA.9}) assuming $q(r)\propto r^{\alpha}(1-r)^{\alpha}$. The three dotted line segments are power functions whose exponents are $0.0$, $1.0$, and $2.0$. The colored lines are the numerical solutions using the different sets of $(\gamma,\alpha)$. The differences of $(\gamma,\alpha)$ are described by the line type for $\gamma$, and the color for $\alpha$.\label{fig:msol}} 
\end{figure}

We assume the intensity function to be a power function $\lambda_{s} = (s\slash a)^{\gamma}\slash\tau$ for a nonzero $\gamma$, where $a$ is a characteristic scale of the fragment size.
In this case, the master equation becomes 
\begin{equation}\label{eqA.7}
\tau \frac{\partial f_{t}(s)}{\partial t} = -\left(\frac{s}{a}\right)^{\gamma}f_{t}(s) + \int_{s}^{\infty}\frac{ds'}{s'}q\left(\frac{s}{s'}\right)\left(\frac{s'}{a}\right)^{\gamma} f_{t}(s').
\end{equation}
As discussed in Ref.~\cite{ito2014-2}, the solution of this master equation has the dynamical scaling property, and the expectation of $s$ obeys
\begin{equation}\label{eqA.8}
\left<s\right> = b t^{-1/\gamma},
\end{equation}
asymptotically, where $b$ is a constant.
Scaling of the fragment size $s$ in Eq.~\eqref{eqA.7} by this $\left<s\right>$ and then ignoring the time-derivative term yield the following asymptotic form 
\begin{equation}\label{eqA.9}
\frac{1}{\gamma}\frac{d \left(xf\right)}{d x}+ \left(\frac{x}{c}\right)^{\gamma} f=
\int_{x}^{\infty}\frac{dx'}{x'}q\left(\frac{x}{x'}\right)\left(\frac{x'}{c}\right)^{\gamma} f(x'),
\end{equation}
where the scale parameter $c\;(=ab)$ is determined so that $\int_{0}^{\infty}dx'\; x'f(x')=1$.
We wish to evaluate the asymptotic functional form of $f(x)$ that follows Eq.~\eqref{eqA.9} in the neighborhood of the origin and at a sufficiently large $x$, because obtaining the full form of $f(x)$ explicitly is difficult in general.
Assuming that the right-hand side of Eq.~\eqref{eqA.9} vanishes provides us with the functional form of the tail (i.e., $f(x)$ at $x\rightarrow\infty$). The leading term is obtained by $f(x)\sim \exp\left[-\left(x\slash c\right)^{\gamma}\right]$.
This stretched-exponential behavior of $f(x)$ itself is robust to $q(r)$; however, the magnitude of the scale parameter $c$ depends on the functional form of $q(r)$.
Estimating the functional form in the neighborhood of the origin by an analytical calculation is generally difficult. 
However, the numerical calculation of Eq.~\eqref{eqA.9} suggests that the functional form in the neighborhood of the origin is characterized by the functional form of $q(r)$.
As shown in Fig.~\ref{fig:msol}, the numerical solution of Eq.~\eqref{eqA.9}, assuming $q(r)$ to be a beta density (i.e., $q(r)\propto r^{\alpha}(1-r)^{\alpha}$), yields $f(x)\sim x^{\alpha}$ in the neighborhood of the origin. This result is independent of the exponent $\gamma$ of the intensity function.

\end{document}